\documentclass[twocolumn]{aastex63}

\usepackage{color,hyperref}

\begin{document}

\title{The Second and Third Data Releases from the UKIRT Hemisphere Survey}

\correspondingauthor{Adam C. Schneider}
\email{adam.c.schneider4.civ@us.navy.mil}

\author[0000-0002-6294-5937]{Adam C. Schneider}
\affil{United States Naval Observatory, Flagstaff Station, 10391 West Naval Observatory Rd., Flagstaff, AZ 86005, USA}

\author{Frederick J. Vrba}
\affil{United States Naval Observatory, Flagstaff Station, 10391 West Naval Observatory Rd., Flagstaff, AZ 86005, USA (retired)}

\author[0000-0002-3858-1205]{Justice Bruursema}
\affil{United States Naval Observatory, Flagstaff Station, 10391 West Naval Observatory Rd., Flagstaff, AZ 86005, USA}

\author[0000-0002-4603-4834]{Jeffrey A. Munn}
\affil{United States Naval Observatory, Flagstaff Station, 10391 West Naval Observatory Rd., Flagstaff, AZ 86005, USA}

\author{Mike Irwin}
\affil{Institute of Astronomy, University of Cambridge, Madingley Road, Cambridge CB3 0HA, UK}

\author{Mike Read}
\affil{Royal Observatory Edinburgh, Blackford Hill, Edinburgh, EH9 3HJ, UK}

\author[0000-0002-6349-7590]{Watson Varricatt}
\affil{UKIRT Observatory, Institute for Astronomy, 640 N. A’ohoku Place, University Park, Hilo, Hawai’i 96720, USA}

\author{Tom Kerr}
\affil{UKIRT Observatory, Institute for Astronomy, 640 N. A’ohoku Place, University Park, Hilo, Hawai’i 96720, USA}

\author[0000-0003-0786-2140]{Klaus Hodapp}
\affil{University of Hawaii, Hilo, HI 96720, USA}

\author[0000-0002-1318-8343]{Simon Dye}
\affil{School of Physics and Astronomy, University of Nottingham, University Park, Nottingham NG7 2RD, UK}

\author{Stephen J. Williams}
\affil{United States Naval Observatory, Flagstaff Station, 10391 West Naval Observatory Rd., Flagstaff, AZ 86005, USA}

\author{Andrew T. Cenko}
\affil{United States Naval Observatory, Flagstaff Station, 10391 West Naval Observatory Rd., Flagstaff, AZ 86005, USA}

\author{Trudy M. Tilleman}
\affil{United States Naval Observatory, Flagstaff Station, 10391 West Naval Observatory Rd., Flagstaff, AZ 86005, USA}

\author[0000-0002-3529-6054]{Marc A. Murison}
\affil{United States Naval Observatory, Flagstaff Station, 10391 West Naval Observatory Rd., Flagstaff, AZ 86005, USA}

\author[0000-0003-2283-2185]{Barry Rothberg}
\affil{United States Naval Observatory, 3450 Massachusetts Avenue NW, Washington, DC 20392-5420, USA}
\affil{Department of Physics and Astronomy, George Mason University, 4400 University Drive, MSN 3F3, Fairfax, VA 22030, USA}

\author[0000-0002-2968-2418]{Scott Dahm}
\affil{Gemini Observatory/NSF's NOIRLab, 950 N. Cherry Avenue, Tucson, AZ 85719, USA}

\author[0000-0002-5604-5254]{Bryan Dorland}
\affil{Office of the Undersecretary of Defense for Research and Engineering, 1000 Defense Pentagon, Washington, DC 20301, USA}

\author[0000-0002-3134-6093]{Andy Lawrence}
\affil{Institute for Astronomy, University of Edinburgh, Royal Observatory, Blackford Hill, Edinburgh EH9 3HJ, UK}

\author[0000-0001-6965-7789]{Kenneth C. Chambers}
\affil{Institute for Astronomy, University of Hawai’i, 2680 Woodlawn Drive, Honolulu, HI 96822, USA}

\begin{abstract}
This paper describes the second and third data releases (DR2 and DR3, respectively) from the ongoing United Kingdom Infrared Telescope (UKIRT) Hemisphere Survey (UHS). DR2 is primarily comprised of the $K$-band portion of the UHS survey, and was released to the public on June 1, 2023.  DR3 mainly includes the $H$-band portion of the survey, with a public release scheduled for September 2025. The $H$- and $K$-band data releases complement the previous $J$-band data release (DR1) from 2018.  The survey covers approximately 12,700 square degrees between declinations of 0\degr\ and $+$60\degr\ and achieves median 5$\sigma$ point source sensitivities of 19.0 mag and 18.0 mag (Vega) for $H$ and $K$, respectively.  The data releases include images and source catalogs which include $\sim$581 million $H$-band detections and $\sim$461 million $K$-band detections.  DR2 and DR3 also include merged catalogs, created by combining $J$- and $K$-band detections (DR2) and $J$-, $H$-, and $K$-band detections (DR3).  The DR2 merged catalog has a total of $\sim$513 million sources, while the DR3 merged catalog contains $\sim$560 million sources.   
\end{abstract}

\section{Introduction}
\label{sec:intro}

Large area surveys spanning a variety of wavelength regimes, including the optical Sloan Digital Sky Survey (SDSS; \citealt{york2000}) and the Panoramic Survey Telescope and Rapid Response System (Pan-STARRS; \citealt{chambers2016}), the near-infrared Two Micron All-Sky Survey (2MASS; \citealt{skrutskie2006}), UKIRT Infrared Deep Sky Surveys (UKIDSS; \citealt{dye2006, lawrence2007, warren2007}), and VISTA Surveys (VHS; \citealt{mcmahon2013}), and the mid-infrared Wide-field Infrared Survey Explorer (WISE; \citealt{wright2010}), have been essential to many areas of astrophysics.  Future large-area surveys will continue to break new ground, with new facilities such as the Rubin Observatory Legacy Survey of Space and Time (LSST; \citealt{ivezic2019}), the Nancy Grace Roman Space Telescope \citep{spergel2015}, the Euclid Space Telescope \citep{laureijs2011}, and the Spectro-Photometer for the History of the Universe, Epoch of Reionization, and Ices Explorer (SPHEREx; \citealt{dore2014}) beginning or soon to begin operations.      

The UKIRT Hemisphere Survey (UHS) is an ongoing large-area near-infrared northern-hemisphere survey conducted using the Wide Field Camera (WFCAM; \citealt{casali2007}) installed at the United Kingdom Infrared Telescope (UKIRT).  The principal objective of UHS is to provide contiguous northern hemisphere near-infrared coverage between declinations of 0\degr\ and $+$60\degr\ by combining over 12,700 square degrees of new imaging with previously observed UKIDSS surveys \citep{lawrence2007}, including the Large Area Survey (LAS), Galactic Plane Survey (GPS), and Galactic Cluster Survey (GCS).  The $J$-band portion of the UHS was previously published in \cite{dye2018}. The $J$-band survey has already enabled high-impact science from diverse astronomical subfields, from the identification of massive high-z quasars (e.g., \citealt{yang2020, banados2023}), the characterization of accreting white dwarfs (e.g., \citealt{gansicke2019, lai2021}), the study of the nearby substellar population (e.g., \citealt{sanghi2023, kirkpatrick2024}), deep surveys of star clusters (e.g., \citealt{esplin2019, miretroig2019}), and solar system bodies (e.g., \citealt{morrison2024}). 

This paper describes the UHS second and third data releases (DR2 and DR3, respectively).  These UHS data releases are a collaborative effort undertaken by the United States Naval Observatory (USNO), the Institute for Astronomy at the University of Hawaii (IfA), the Cambridge Astronomy Survey Unit (CASU), and the Wide Field Astronomy Unit (WFAU) in Edinburgh. This paper is organized as follows: in Section \ref{sec:survey} we describe the general characteristics of the surveys, including survey design, execution, data acquisition and processing, and quality control (QC). In Section \ref{sec:cat} we describe the available catalogs created by the survey.  The data, which includes catalogs and images, are made publicly available through the WFCAM Science Archive\footnote{http://wsa.roe.ac.uk} (WSA; \citealt{hambly2008}) maintained by the WFAU. We summarize this work in Section \ref{sec:summary} and provide some information regarding ongoing and future UKIRT surveys.    

\section{Survey Overview}
\label{sec:survey}

\begin{figure*}
\plotone{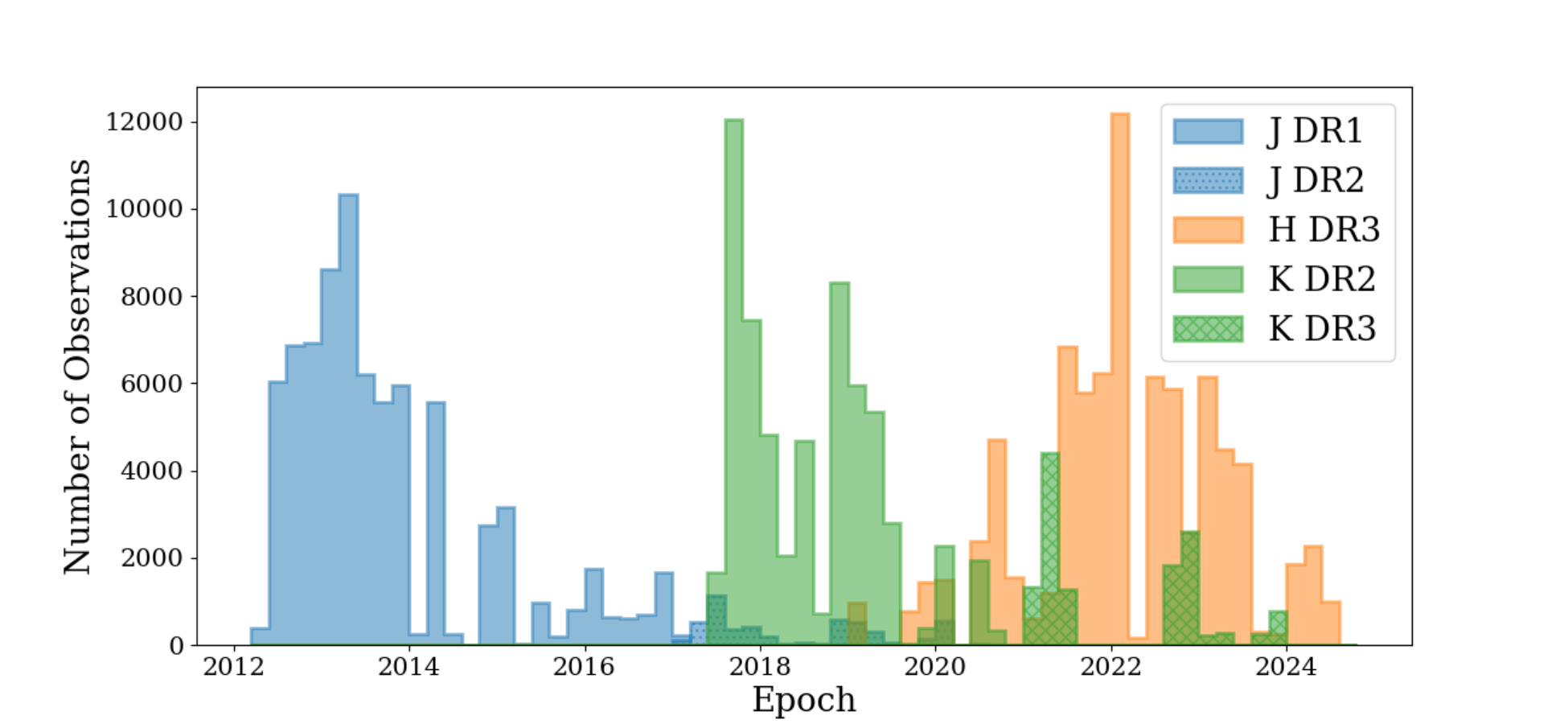}
\caption{The number of observations (Stacks) as a function of date for DR1, DR2, and DR3.  While DR1, DR2, and DR3 generally correspond to $J$-band, $K$-band, and $H$-band observations, respectively, the hatched regions show the dates of additional $J$-band observations included in DR2 and the dates of additional $K$-band observations included in DR3.} 
\label{fig:datehist}
\end{figure*}

\begin{figure*}
\plotone{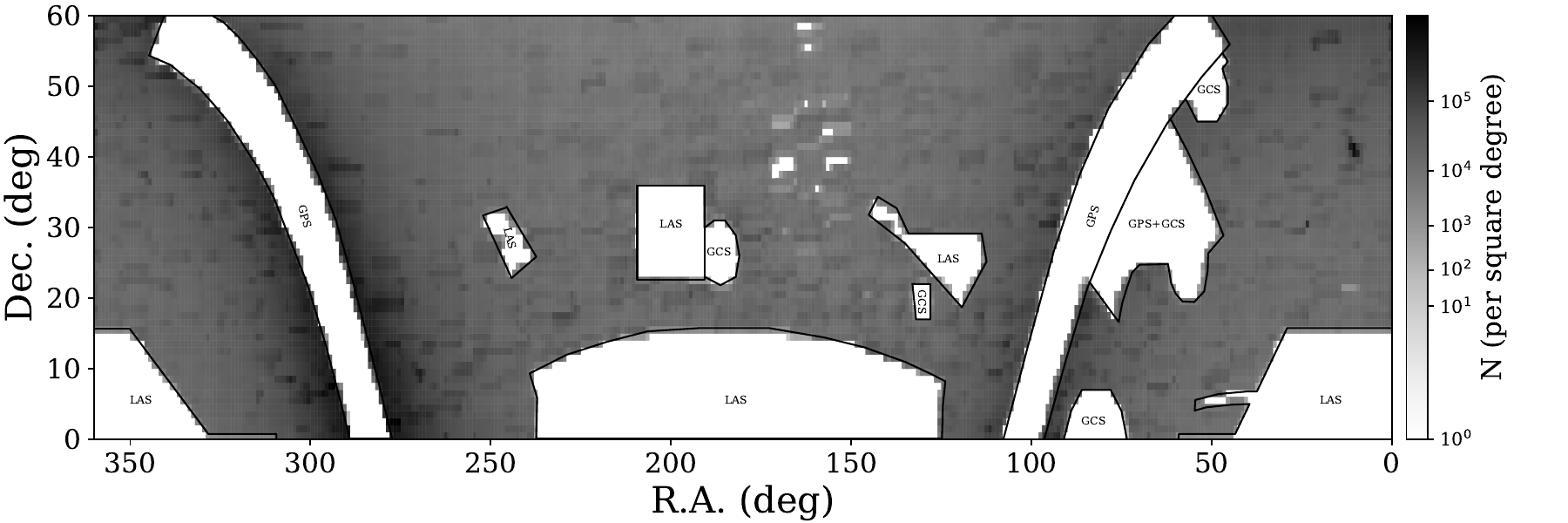}
\caption{UHS $H$-band DR3 coverage map showing the number of detections per square degree.  Areas previously covered by the UKIDSS Large Area Survey (LAS), Galactic Plane Survey (GPS), and Galactic Cluster Survey (GCS) are indicated. } 
\label{fig:hcoverage}
\end{figure*}

\begin{figure*}
\plotone{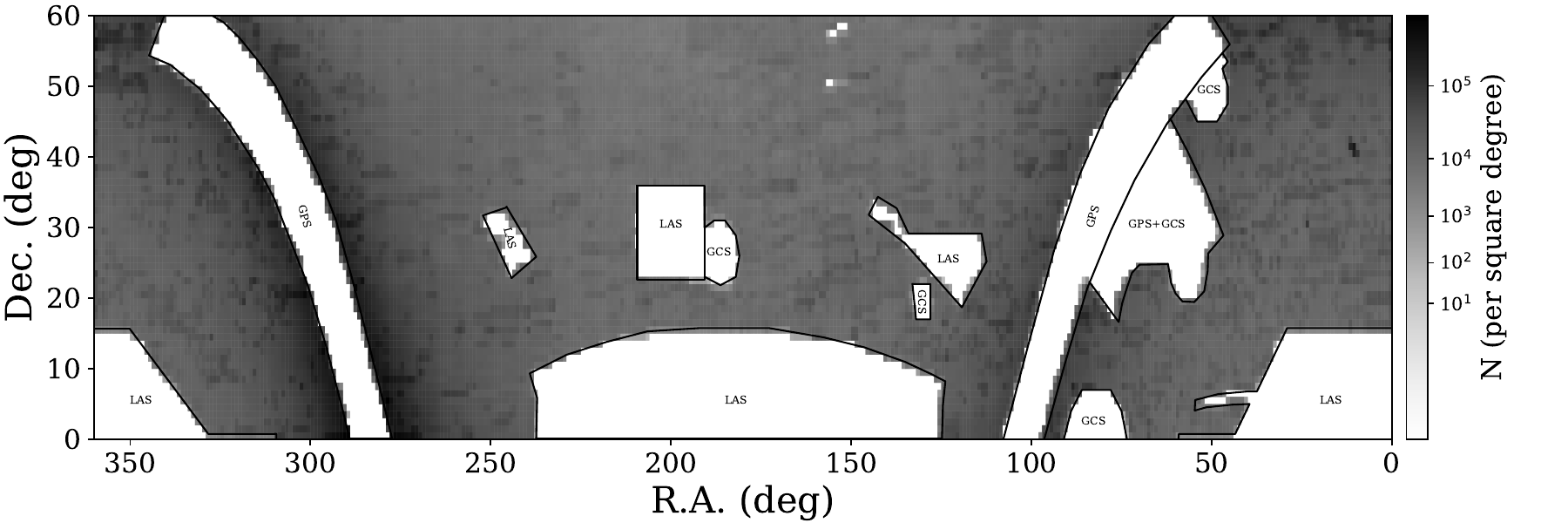}
\caption{UHS $K$-band coverage map showing the number of detections per square degree, which includes all frames from DR2 and DR3.  Areas previously covered by the UKIDSS Large Area Survey (LAS), Galactic Plane Survey (GPS), and Galactic Cluster Survey (GCS) are indicated. }  
\label{fig:kcoverage}
\end{figure*}

DR1 was publicly released on 1 August 2018 \citep{dye2018} and contains the vast majority of the $J$-band portion of the UHS survey ($\sim$98\% of the UHS surveyed area).  The DR2 data release was focused on $K$-band observations, though a small amount of remaining or re-observed $J$-band observations were also included.  Similarly, the DR3 data release is focused on $H$-band observations, but also includes new or re-observed $K$-band data.  Figure \ref{fig:datehist} shows the observation dates for $J$-, $H$-, and $K$-band UHS data, highlighting what dates were included in DR1, DR2, and DR3 releases. $K$-band observations started in July 2017, and the latest $K$-band data included in DR2 occurred in August 2020. $J$-band observations between January 2017 and August 2020 are also included in DR2 and DR3, however, these $J$-band data have not yet fully passed through our QC process.  A detailed description regarding how to access data that have not yet been fully QC'd is given in Section \ref{sec:cat}.  $H$-band observations started in January 2019, and the last data included in DR3 occurred in July 2024.  DR3 also includes all $K$-band images taken between August 2020 and July 2024.  However, as with the $J$-band data in DR2, these $K$-band data have not undergone full QC.  Figures \ref{fig:hcoverage} and \ref{fig:kcoverage} show the coverage and relative completeness of the $H$-band and $K$-band observations (respectively).  The $K$-band coverage map includes all $K$-band data from DR2 and DR3, while the $H$-band coverage map is all $H$-band data from DR3.  The vast majority of the areas uncovered by UHS in the figures are those regions already covered by various UKIDSS surveys.        

The survey strategy and reduction processes for DR1 are discussed in great detail in \cite{dye2018}.  Generally, the same strategy and reduction processes are adopted in DR2 and DR3.  We summarize salient steps here and highlight any significant differences between DR1, DR2, and DR3.

\subsection{Nomenclature}
To describe the surveys, we adopt the same nomenclature as was used for DR1 as well as previous UKIDSS surveys.  Survey-specific terms include {\it Exposure} (a single 10 second WFCAM integration), {\it Stack} (coadd of four Exposures), and {\it Tile} (the contiguous area formed by four Stack positions in a 2$\times$2 grid).  Observations are scheduled using `Minimum Schedulable Blocks' (MSBs), that are typically made up of 32 Stack frames (or 8 Tiles), which have execution times of $\sim$30 minutes. 

\subsection{Survey Design}
The geometry and tiling of UHS observations is described in \cite{dye2018}.  Briefly, the survey area was defined using the Survey Definition Tool, which forms contiguous areas by overlapping tiles by 0.5 arcmin in right ascension (RA) and declination (Dec.). The same RA and Dec coordinates that were used for the $J$-band observations in DR1 are also used for all $H$- and $K$-band observations in DR2 and DR3.  These observations also use the same guide stars.  As with the $J$-band observations in DR1, the $H$- and $K$-band observations are split into 24 projects with each project spanning 1 hour in right ascension.  One additional project was created for both $H$- and $K$-band observations reserved for `patch' observations (i.e., re-observed Stacks that did not pass QC).  These project names appear in the \texttt{Multiframe} table in the WSA.

$H$-band MSBs were permitted to be executed if the full-width half-maximum (FWHM) of the point spread function (PSF) was less than 1\arcsec\ and conditions at the UKIRT site were photometric.  $K$-band MSBs could be executed if the FWHM of the PSF was less than 1\farcs1 in photometric conditions.  No $J$-band sky brightness limits were imposed for $H$- or $K$-band observations.  Each $K$-band observation is made up of four 10 s Exposures, which are combined into a single Stack frame.  While the $K$-band survey was modeled after the $J$-band survey, using 10 s exposures, the $H$-band survey was changed to have observations made up of 4 Exposures consisting of 2 5-second coadds in order to keep background levels low.  No microstepping is done for either the $H$- or $K$-band observations, consistent with previous UHS $J$-band observations.  Thus, the $J$-, $H$-, and $K$-band UHS images are limited by the angular size of the WFCAM pixels (0\farcs4) and may not be Nyquist sampled in all cases.

\subsection{Data Processing and Analysis}

Following UKIDSS and UHS DR1, the UHS DR2 and DR3 data are processed with the CASU-developed VISTA/WFCAM data reduction pipeline (DRP; \citealt{irwin2004, hambly2008}).  All UHS data are initially transferred to CASU, where the raw frames are flattened, sky-subtracted, and stacked.  Intermediate catalogs are then generated for each stacked frame, while final merged catalogs are generated in a later step (see Section \ref{sec:cat}).

\subsubsection{Astrometric Calibration}
UHS is astrometrically calibrated against 2MASS \citep{skrutskie2006}, as described in \cite{irwin2004}.  As discussed in \cite{dye2018}, UHS astrometry is correlated with Galactic latitude, as the number of bright 2MASS calibration sources decreases with higher latitudes, leading to a decrease in astrometric precision.  A comparison of the DR2 and DR3 astrometric solutions to objects in common with Gaia DR3 \citep{gaia2023} suggests a typical positional accuracy of $\sim$50 mas.

Some early success recalibrating the astrometric solution for UHS against Gaia can be found in in \cite{schneider2023} and \cite{schneider2024}. Astrometric recalibration of all UHS (and UKIDSS) catalogs against Gaia would result in significant improvements over the current astrometric solutions.  Initial tests show that median Gaia positional offsets could be reduced from $\sim$50 mas to $\sim$8 mas.   

\subsubsection{Photometric Calibration}
All UHS observations use the UKIRT/WFCAM photometric system \citep{hewett2006}.  Photometric calibration against 2MASS sources is carried out by the DRP at CASU using the methods detailed in \cite{hodgkin2009}.  For each detector, sources are cross-matched with 2MASS using a 1\arcsec\ matching radius.  Accounting for astrometric distortion and 2MASS to WFCAM filter differences, most frames are calibrated using 25 to 1000 2MASS sources, achieving a photometric accuracy of approximately 2\%.

\subsubsection{Quality Control}
\label{sec:data}

\begin{figure*}
\plotone{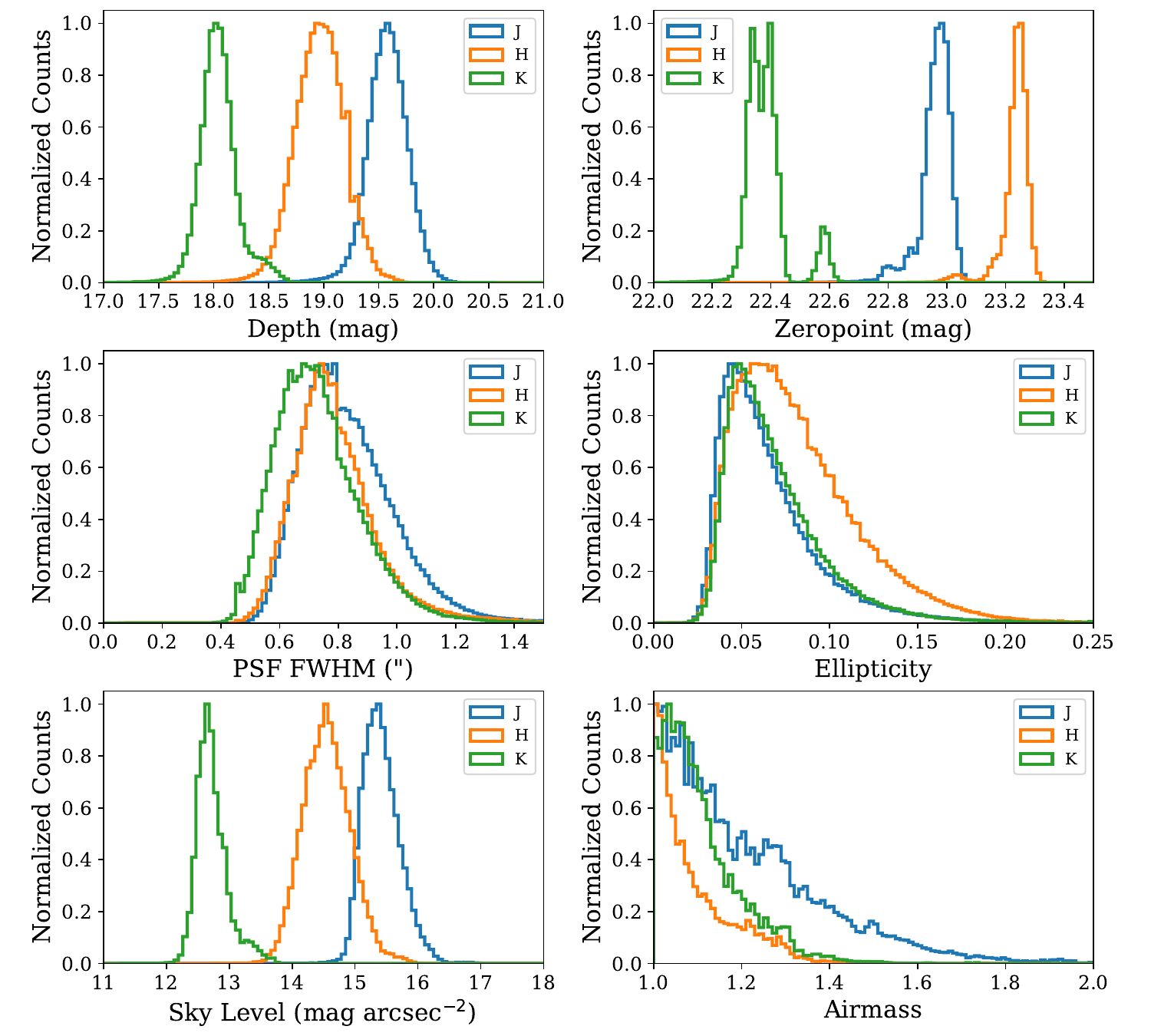}
\caption{Data quality measurements for all $J$-, $H$-, and $K$-band UHS frames, including depth (Vega magnitudes), PSF FWHM (arcsec), stellar ellipticity, zero-point magnitude, sky brightness (mag arcsec$^{-2}$) and airmass.} 
\label{fig:numbers}
\end{figure*}

After the data are processed at CASU with the DRP, frames are transferred to USNO for additional QC.  The first QC step is to inspect the metrics produced by the DRP for each Stack frame to ensure that they satisfy the requirements set out for each filter.  For  UHS $H$- and $K$-band observations, frames are required to have measured seeing $\leq$ 1\farcs1 and stellar ellipticity $\leq$0.2.  $K$-band frames are also required to have 5$\sigma$ limiting magnitudes $\geq$ 17.6 mag, while $H$-band frames are required to have 5$\sigma$ limiting magnitudes $\geq$ 18.4 mag.  Stack frames that do not meet these requirements are flagged and placed into a patch program for re-observation.  Figure \ref{fig:numbers} displays the quality metrics for $H$- and $K$-band UHS observations, compared to previously published metrics for $J$-band observations from \cite{dye2018}.  

After image metrics are evaluated, USNO then visually inspects all of the remaining processed image Stacks for defects or problems that otherwise were not identified by the automated CASU processing.  The most common causes of flagged frames during the visual inspection phase are trailed images or issues with sky subtraction. Frames identified as problematic are also placed into a patch program for re-observation.

\subsection{Degradation of Array \#4}
\label{sec:array4}
\begin{figure*}
\plotone{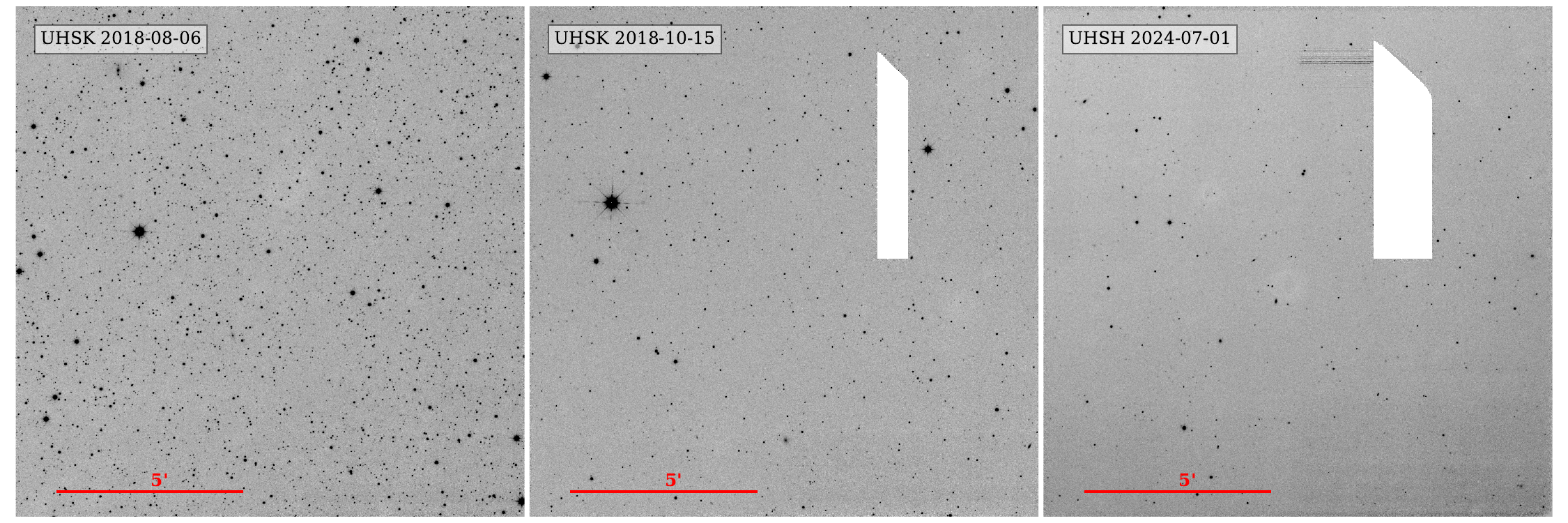}
\caption{WFCAM array \#4 before (left) and after (center) the appearance of dead pixels described in Section \ref{sec:array4}.  The right image shows the size of the dead pixel region for the last $H$-band image included in DR3.} 
\label{fig:gap}
\end{figure*}

Each WFCAM detector is divided into four quadrants, and each quadrant is divided into eight different channels (see \citealt{dye2018}, Figure 1). Beginning in 2017 during $K$-band (DR2) survey observations, one of the channels on detector \#4 began to degrade after WFCAM warmed up following a power outage at UKIRT.  This was initially seen as a region of low-count features spanning $\sim$15-20 columns of the detector.  A subsequent warm-up of WFCAM in August 2018 revealed a dead pixel area on on detector \#4 that has continued to grow with each detector thermal cycle.  Figure \ref{fig:gap} shows the last UHS $K$-band image before the pixels in the affected region completely died (6 August 2018), and the first appearance of the affected region in a UHS $K$-band image (15 October 2018).  The third panel in the figure shows the current size of the affected region on the latest $H$-band image included in the DR3 data release. The total affected area encompasses $\sim$6\% of the \#4 detector, or $\sim$1.5\% of the total area of all four detectors.  Because only a small portion of detector \#4 was affected, no further effort was undertaken to account for and cover this gap in the detector with additional survey images. Thus, some small fraction (1--2\%) of the sky in nominally observed areas in the UHS $H$- and $K$-band portions of the survey will not have image or detection data available.       

\section{Catalogs}
\label{sec:cat}

\subsection{Images and Detections}
DR1 contained a total of 76,355 $J$-band Stack frames.  Of these frames, 11,140 were flagged as deprecated, with reasons including being re-observed, not passing QC metrics, or not passing by-eye image checks.  There are thus 65,215 ``good'' $J$-band frames in DR1.  DR2 contains all DR1 $J$-band data plus 5,026 additional $J$-band Stack frames, while no new $J$-band data are introduced between DR2 and DR3.  The images and detection tables for the new DR2 $J$-band frames are available through the WSA, however these images have not yet undergone full QC evaluation.  QC for these frames is underway and the results will be included in a future UHS data release.  Images that have not yet undergone full QC taken before July 2024 can still be accessed through the WSA in the same manner as fully QC'd images (e.g., GetImage or Freeform SQL).  Such images will have a numeric code of `85' in the `deprecated' column of their corresponding multiframeID.  Position and photometric measurements from these non-QC'd images are not available in the commonly used WSA catalog search tools (e.g., Region Search or CrossID), as these tools only query the \texttt{uhsDetection} or \texttt{uhsSource} tables.  Detections from the non-QC'd images are present in the \texttt{uhsDetectionAll} tables, which is accessible through the Freeform SQL search function.  Detailed information regarding the use of the Freeform SQL tool can be found in \cite{hambly2008} and the WSA SQL Cookbook\footnote{http://wsa.roe.ac.uk/sqlcookbook.html}.  The number of $J$-band detections  therefore remains the same in the \texttt{uhsDetection} table for DR1 and DR2, but the number of entries increases for the \texttt{uhsDetectionAll} table between DR1 and DR2.  There are $\sim$481 million $J$-band detections in the \texttt{uhsDetection} tables in DR1, DR2, and DR3, while there are $\sim$563 million $J$-band detections in the DR1 \texttt{uhsDetectionAll} table and $\sim$582 million $J$-band detections in the DR2 and DR3 \texttt{uhsDetectionAll} tables.   

DR2 contains a total of 66,341 $K$-band Stack frames, 12,371 of which have been flagged as deprecated, leaving 53,970 ``good'' $K$-band frames.  DR3 contains an additional 13,336 $K$-band frames.  These 13,336 $K$-band frames have not been fully QC'd.  There are $\sim$339 million $K$-band detections in the DR2 \texttt{uhsDetection} table and $\sim$393 million $K$-band detections in the DR2 \texttt{uhsDetectionAll} table.  The DR2 and DR3 \texttt{uhsDetection} tables contain the same number of $K$-band detections, while the DR3 \texttt{uhsDetectionAll} table contains $\sim$462 million $K$-band detections.  

DR3 includes 81,462 $H$-band Stack frames, of which 26,512 have been flagged as deprecated, leaving 54,950 ``good'' $H$-band frames.  There are $\sim$439 million $H$-band detections in the DR3 \texttt{uhsDetection} table and $\sim$581 million $H$-band detections in the DR3 \texttt{uhsDetectionAll} table. A summary of the number of detections for each filter for each UHS data release is provided in Table \ref{tab:jhknumbers}.

\begin{deluxetable*}{lccccccccc}
\tablecaption{UHS Detection Summary}
\label{tab:jhknumbers}
\tablehead{\colhead{Table/Filter} & \colhead{DR1} & \colhead{DR2} & \colhead{DR3}\\
\colhead{} & \colhead{\# of detections} & \colhead{\# of detections} & \colhead{\# of detections}} 
\startdata
\texttt{uhsDetection} ($J$) & 481,056,015 & 481,056,015 & 481,056,015 \\
\texttt{uhsDetectionAll} ($J$) & 562,887,595 & 581,620,056 & 581,620,056 \\
\texttt{uhsDetection} ($K$) & \dots & 338,996,723 & 338,996,723 \\
\texttt{uhsDetectionAll} ($K$) & \dots & 392,676,109 & 461,795,508 \\
\texttt{uhsDetection} ($H$) & \dots & \dots & 438,720,157 \\
\texttt{uhsDetectionAll} ($H$) & \dots & \dots & 581,089,560 \\
\enddata
\end{deluxetable*}

\subsection{Merged Catalogs}
Processed images and detection tables are transferred to WFAU to create source catalogs for each data release. Source catalogs are created by combining duplicate sources in the same filter on overlapping fields and matching positions among the available filters for merged catalogs.  The source-merging process follows the methods described in \cite{hambly2008}. Only detections from the ``good'' Stack frames (e.g., the \texttt{uhsDetection} tables) are included in the final merged catalogs.  The matching radius for merging sources in the DR2 and DR3 merged catalogs is 2\arcsec.  As described in \cite{hambly2008}, there are cases where the pairing algorithm does not correctly match detections, such as high proper motion objects (depending on the total proper motion of the object and the time between observations).  However, non-matched sources are still included in the merged catalogs so all information for a source that was not correctly paired by the algorithm can still be recovered.  

The DR2 merged catalog (\texttt{uhsSource}) contains $\sim$513 million sources, $\sim$306 million of which have both $J$- and $K$-band detections. The DR3 merged catalog has $\sim$560 million sources, where $\sim$268 million have $J$-, $H$-, and $K$-band detections.

For DR3, proper motions are also provided for each merged source.  Proper motions are calculated following the methods described in \cite{collins2012}. The epoch tolerance (e.g., the minimum time between the maximum and minimum epochs) for UHS proper motions is set to 550 days, which should be much smaller than the time between observations for objects detected with more than one filter (see Figure \ref{fig:datehist}).

\subsection{Basic UHS Data Access}

Here we describe how to carry out simple UHS queries through the WSA for finding images and catalog entries for a single coordinate search.  After navigating to the WSA landing page (http://wsa.roe.ac.uk/), available UHS images for a single position can be found by selecting ``GetImage'' from the available links on the left side of the page.  On the following Image Extraction page, there is a drop-down menu for selecting the Database release to use where UHS DR1, DR2, or DR3 can be chosen. A similar drop-down menu exists for the programme/survey to use, and for all UHS data releases the only usable option from this menu is the UKIRT Hemisphere Survey, UHS.  Once these options have been selected, simply enter the desired coordinates for the search and any available images will be returned.  

A basic UHS catalog search can be found by selecting ``Region'' from the left-hand menu on the WSA landing page.  Similar to the simple image search, UHS DR1, DR2, or DR3 can be selected from the drop-down menu for the Database release to use, while UKIRT Hemisphere Survey, UHS should be chosen from the programme/survey drop-down menu.  One additional drop-down menu exists for the region search that allows a user to specify between a search of the detection table (unmerged) or the source table (merged).  Coordinates and a search radius (up to 90\arcmin) can then be entered, with the results provided as an HTML table or saved as ascii, fits, or VOTable files.

\section{Summary and Future Work}
\label{sec:summary}

With the release of DR2 and DR3, there is now near-complete $J$-, $H$-, and $K$-band coverage over the UHS survey area, with survey depths of 19.6 mag, 19.0 mag, and 18.0 mag, respectively. The UHS survey combined with previous UKIDSS surveys now provides almost-complete near-infrared coverage in the northern hemisphere between 0\degr\ and $+$60\degr. UKIRT employs an English Equatorial mount, which has a hard mechanical limit of $+$60\degr\ for pointing the telescope. In the southern hemisphere, surveys carried out using the Visible and Infrared Survey Telescope for Astronomy (VISTA; \citealt{sutherland2015}) provide similar depth in the near-infrared, including the VISTA Hemisphere Survey (VHS; \citealt{mcmahon2013}),  the VISTA Variables in the Via Lactea survey (VVV; \citealt{saito2012}), and the VISTA Kilo-degree Infrared Galaxy survey (VIKING; \citealt{edge2013}).  The only area of the sky therefore that has not been covered by UKIRT or VISTA surveys is $\delta >$$+$60\degr.

Current UKIRT survey observations are focused on any remaining UHS $H$- and $K$-band MSBs that have not yet been observed as well as observations from the $H$- and $K$-band patch projects.  UKIRT is also extending the wavelength coverage of the UHS survey area with $Y$-band observations while simultaneously providing second-epoch $J$-band observations.  The second-epoch $J$-band observations will provide a time baseline of $\sim$10 years at a single wavelength for studies where changes in positions (e.g., proper motions) or brightness (e.g., long-term variability) are beneficial. 

\section{Acknowledgements}

This publication makes use of data products from the UKIRT Hemisphere Survey, which is a joint project of the United States Naval Observatory, the University of Hawaii Institute for Astronomy, the Cambridge University Cambridge Astronomy Survey Unit, and the University of Edinburgh Wide-Field Astronomy Unit (WFAU). The WFAU gratefully acknowledges support for this work from the Science and Technology Facilities Council (STFC) through ST/T002956/1 and previous grants. SD acknowledges support from a UK STFC grant (ST/X000982/1). This publication makes use of data products from the Two Micron All Sky Survey, which is a joint project of the University of Massachusetts and the Infrared Processing and Analysis Center/California Institute of Technology, funded by the National Aeronautics and Space Administration and the National Science Foundation. The authors wish to recognize and acknowledge the very significant cultural role and reverence that the summit of Mauna Kea has always had within the indigenous Hawaiian community. We are extremely grateful to have the opportunity to conduct observations from this mountain. The authors also sincerely thank the UKIRT staff for carrying out the UHS observations.

\clearpage
\bibliography{references}{}
\bibliographystyle{aasjournal}

\end{document}